\documentclass[preprints,article,accept,oneauthor]{Definitions/mdpi}
\firstpage{1}  
 \pubvolume{1} \issuenum{1} \articlenumber{0}
\pubyear{2023} \copyrightyear{2023}
\datereceived{ }
\daterevised{ }
\dateaccepted{ }
\datepublished{ }
\hreflink{https://doi.org/}
\pdfoutput=1

\Title{Mission Target: Tetraquark Mesons of Flavour-Cryptoexotic Type}
\TitleCitation{Mission Target: Tetraquark Mesons of Flavour-Cryptoexotic Type}
\Author{Wolfgang Lucha}
\AuthorNames{Wolfgang Lucha}
\AuthorCitation{Lucha, W.}
\address[1]{Institute for High Energy Physics, Austrian Academy of Sciences, Nikolsdorfergasse 18, A-1050 Vienna, Austria; {wolfgang.lucha@oeaw.ac.at}}
\corres{Correspondence: Wolfgang.Lucha@oeaw.ac.at (W.L.)}
\abstract{Currently, flavour-cryptoexotic tetraquarks form the most common sort of all experimentally established exotic multiquark hadrons. This note points out a few promising concepts that should help improve theoretical (but, for several reasons, not quite straightforward) analyses of such kind~of~states: among others, their scope of application encompasses the strong interactions in the limit of (arbitrarily) large numbers of colours, and equally analytical and nonperturbative approaches to multiquark~states.}
\begin{document}

\section{Species of Bound States Within Quantum Chromodynamics: The Crucial Diverseness}\label{sE}Within the framework of theoretical elementary particle physics, the strong interactions are described, at the \emph{presently} most fundamental level, by quantum chromodynamics (QCD), a renormalizable relativistic \emph{quantum field theory} invariant under local gauge transformations constituting representations of SU(3), the non-Abelian, compact, special unitary Lie group of degree three. Presumably out of sheer desperation, the degree of freedom introduced by this gauge symmetry has been called \emph{colour}; that naming is reflected by the related quantum~field
theory, QCD. The two sorts of dynamical degrees of freedom of QCD transform according to irreducible representations of SU(3): the massless (spin-$1$) vector gauge bosons called~gluons necessarily according to the eight-dimensional adjoint representation of SU(3), and any~of its spin-$\frac{1}{2}$ fermions subsumed under the notion of ``quarks'' according to the three-dimensional fundamental representation of SU(3). The different members of the set of quarks, henceforth denoted by $q_a$, are discriminated by their \emph{flavour} quantum number, $a\in\{u,d,s,c,b,t(,\dots?)\}$. Consequently, the strong coupling $g_{\rm s}$, governing both the interactions of three or four~gluons among each other and the couplings of gluons to quarks, and, because of its usually~pairwise occurrence, frequently entering analyses under the disguise of a strong \emph{fine-structure~coupling}\begin{equation}\alpha_{\rm s}\equiv\frac{g_{\rm s}^2}{4\pi}\ ,\label{as}\end{equation}and the masses $m_a$ of the quarks $q_a$ constitute the \emph{totality} of fundamental parameters~of~QCD.

The mathematical structure of QCD as a \emph{gauge} theory is, comparatively, simple: an even unbroken gauge symmetry based on a simple albeit non-Abelian group. Nevertheless, QCD produces various peculiar phenomena. Most prominent among these is the \emph{confinement} of all colour degrees of freedom: The \emph{dynamical} degrees of freedom of QCD, its quarks and gluons, each carrying some definitely nonzero amount of colour, are not experimentally~accessible in an isolated form. In fact, \emph{exclusively} those of their \emph{bound states} that transform under the gauge group as singlets, collectively subsumed under the name of \emph{hadrons}, are directly~observable.\footnote{\emph{Group-theoretically}, the decomposition of the tensor product of the SU(3) representations of \emph{all} constituents of any hadron into \emph{irreducible} SU(3) representations must include one (one-dimensional) singlet representation,~at~least.} On conceptual grounds, these hadron states are to be separated into two disjoint \emph{subdivisions}:\begin{itemize}\item The ordinary quark--antiquark mesons and three-quark baryons are called \emph{conventional}.\item All other (hence non-conventional) types of hadrons --- multiquark states, quark--gluon hybrid mesons, totally gluonic glueballs --- are captured by the notion of~\emph{exotic}~hadrons.\end{itemize}

Needless to stress, it would be desirable if the \emph{suspected} (exotic) multiquark nature of an experimentally detected state could be straightforwardly decided already from its \emph{apparently} non-conventional number of different quark flavours. Such fortunate coincidence,~however, may only happen if, for strictly none of the quarks and antiquarks constituting the candidate multiquark state under investigation, its specific flavour gets \emph{counterbalanced} by the \emph{respective} antiflavour. Otherwise, the multiquark will not be able to exhibit this particular flavour. The precise identification of all the states where such compensations do take place is~provided~by\begin{Definition}\label{dFC}A multiquark hadron is termed \textbf{flavour-cryptoexotic} if it does not exhibit more~open quark flavours than the corresponding category of conventional hadrons, that is, trivially, at~most~two open quark flavours in the case of meson states or three open quark flavours in the case of~baryon~states.\end{Definition}

Now, for every multiquark exotic hadron of the flavour-cryptoexotic type its net overall content of quark flavour not counterbalanced by an antiflavour is, by Definition~\ref{dFC}, \emph{necessarily} identical to that of its certainly existing counterpart(s) among the set of \emph{conventional} hadrons. As a consequence thereof, in the investigation of \emph{flavour-cryptoexotic} hadron states there~arise (additional) complications that are not encountered in any analysis of multiquark states~built from quarks and antiquarks carrying totally disparate quark flavours. Namely, for matching spin and parity quantum numbers any flavour-cryptoexotic hadron can and, quite~generally, will undergo \emph{mixing} with its appropriate conventional counterpart; this circumstance~should be taken into account in comprehensive descriptions of the \emph{former} hadron (cf., e.g., Section~\ref{sN}).

The category of multiquark states with the \emph{least} number of constituents is formed by the totality of all \emph{tetraquark} mesons, generically identified by $T$: bound states of two quarks~$q_b,q_d$ and two antiquarks $\overline q_a,\overline q_c$ with flavour quantum numbers $a,b,c,d\in\{u,d,s,c,b\}$ and masses $m_a,m_b,m_c,m_d$. Utilizing Definition~\ref{dFC}, let us put our focus on their \emph{flavour-cryptoexotic} variant\begin{equation}T=[\overline q_a\,q_b\,\overline q_b\,q_c]\ ,\qquad a,b,c\in\{u,d,s,c,b\}\ .\label{fce}\end{equation}The (presumably not surprising) {motivation} for this choice is predicated on the circumstance that the subset of flavour-cryptoexotic tetraquarks (\ref{fce}) constitutes that sort of multiquarks for which (at present) the largest count of \emph{candidate hadrons} has been observed by experiment~\cite{PDG}. In order to pave the way towards (eventually optimized) approaches to flavour-cryptoexotic tetraquarks, it appears to be worthwhile to collect, in a systematic manner, various concepts, notions, or pieces of information, proposed and ``lurking around'' \cite{LMS-N1,LMS-N2,LMS-N3,LMS-N3+,LMS-N3.1,LMS-N4,LMS-N7} (Sections~\ref{sC}, \ref{sN}, and~\ref{sS}).

Table~\ref{TC} categorizes the \emph{conceivable} configurations of all flavour-cryptoexotic tetraquarks. Simplicity suggests to adopt, \emph{for illustration}, states that involve \emph{three} mutually different~quark flavours, i.e., two \emph{open} flavours and a compensation of a different flavour and its antiflavour:\begin{equation}T=[\overline q_a\,q_b\,\overline q_b\,q_c]\ ,\qquad a,b,c\in\{u,d,s,c,b\}\ ,\qquad a\ne b\ne c\ .\label{3F}\end{equation}

\begin{table}[H]\caption{Flavour-\emph{cryptoexotic} tetraquarks (\ref{fce}) distinguished by different and open quark-flavour~content $a\ne b\ne c$: \emph{open flavour} means any flavour not compensated by its antiflavour; distilled from~Refs.~\cite{LMS-N4,L:MT1}.\label{TC}}\begin{center}\begin{tabular}{ccc}\toprule\textbf{Number of Different}&$\quad$ \ \textbf{Quark Composition} \ $\quad$&\textbf{Number of Open}\\\textbf{Quark Flavours Involved}&$\overline q_\square\,q_\square\ \,\overline q_\square\,q_\square$&\textbf{Quark Flavours Involved}\\\midrule 3&$\overline q_a\,q_b\ \,\overline q_b\,q_c$&2\\&$\overline q_a\,q_b\ \,\overline q_c\,q_c$&2\\\midrule
2&$\overline q_a\,q_a\ \,\overline q_a\,q_b$&2\\&$\overline q_a\,q_a\ \,\overline q_b\,q_a$&2\\&$\overline q_a\,q_b\ \,\overline q_b\,q_a$&0\\&$\overline q_a\,q_a\ \,\overline q_b\,q_b$&0\\\midrule 1&$\overline q_a\,q_a\ \,\overline q_a\,q_a$&0\\\bottomrule\end{tabular}\end{center}\end{table}

\section{\emph{Multiquark-Phile} Four-Point Correlation Functions of Hadron Interpolating Operators}\label{sC}Within the framework of QCD, productive contact between the coloured (thus confined and experimentally not \emph{directly} accessible) dynamical degrees of freedom of QCD, on~the~one hand, and the colourless (hence experimentally observable) hadrons, on the other hand,~may be established by means of the tool labelled hadron interpolating operators. By construction, (a candidate for) any such \emph{hadron interpolating operator} is an (inevitably gauge-invariant)~local operator, defined in terms of the quark and gluon fields, carrying the same flavour, spin, and parity quantum numbers as the hadron under consideration, and getting its overlap~with the hadron testified by its \emph{nonzero matrix element} between the hadron state and the QCD~vacuum.

For the subsequent discussions, neither spin nor parity degrees of freedom appear to be of utmost importance. Thus, upon suppressing Dirac matrices and Lorentz structures, in the following all reference to spin and parity in a hadron interpolating operator may be~dropped. For the interpolating operator of a \emph{conventional} meson composed of an antiquark of flavour $a$ and a quark of flavour $b$, the (generic) quark--antiquark bilinear-current form may be~chosen:\begin{equation}j_{\overline ab}(x)\equiv\overline q_a(x)\,q_b(x)\ .\label{b}\end{equation}

Within quantum field theories, information about bound states of some basic degrees of freedom may be deduced from the contributions of the bound states, in form of intermediate states, to scattering processes of the \emph{potential} bound-state constituents. The related scattering amplitudes encoding this information can be derived from appropriate correlation functions of the particles involved in these scattering reactions. An ($n$-point) {correlation function} is, by definition, the vacuum expectation value of the (time-ordered) product of the field~operators of the ($n=2,3,\dots$) particles under consideration. (Below, the time-ordering \emph{prescription} will be indicated by the symbol T, whereas the respective vacuum states are notationally reduced to their mere state brackets.) Upon truncation of the propagators of all external~particles,~any scattering amplitudes aimed at result as the on-mass-shell limits of the correlation functions.

For its application to the tetraquark mesons, the (standard) procedure for the extraction of bound-state properties sketched above necessitates an inspection of scattering amplitudes formalizing scattering processes of two conventional mesons into two conventional mesons. Scattering amplitudes for reactions of this kind may receive intermediate-state contributions of some tetraquark mesons. These bound states would reveal their existence by contributing in form of intermediate-state poles. Now, adhering to the outlined procedures, the scattering amplitudes can be derived from the associated four-point correlation functions. In these, the conventional mesons undergoing scattering may enter (from the aspect of \emph{QCD}) by means of the interpolating currents (\ref{b}). Consequently, for a moment leaving aside all details related to the involved quark flavours $a,b,c,d$, approaching the features of tetraquark mesons~requires coping with \emph{generic} four-point correlation functions of quark--antiquark~bilinear currents~(\ref{b}),\begin{equation}\left\langle{\rm T}\!\left(j(y)\,j(y')\,j^\dag(x)\,j^\dag(x')\right)\right\rangle\ .\label{4}\end{equation}

Evidently, the first and primary goal of all of the analyses envisaged above has to be the both unique and unambiguous identification of those contributions to four-point correlation functions (\ref{4}) that might support the development of an intermediate-state pole~interpretable as being related to an exotic tetraquark state. This task may be accomplished by, for instance, identifying and, subsequently, discarding all \emph{those} contributions of QCD origin that certainly exhibit no relationship to any \emph{tetraquark} state. To this end, by use of the Mandelstam variable\begin{equation}s\equiv(p_1+p_2)^2=(q_1+q_2)^2\label{Mv}\end{equation}(fixed by either the external momenta of the initial state $p_1,p_2$ or the external momenta of the final state $q_1,q_2$) a simple but \emph{rigorous}, maybe even useful \emph{criterion} \cite{LMS-N1,LMS-N3} has been~formulated:\begin{Proposition}\label{dpTP}A QCD-level contribution to a four-point correlation function (\ref{4}) is considered to be potentially capable of supporting (the formation of) intermediate-state tetraquark poles and~referred to as \textbf{tetraquark-phile} \cite{LMS-N2,LMS-N5} if, as a function of the Mandelstam variable $s$, it exhibits a nonpolynomial dependence on $s$ and develops a four-quark intermediate-state \emph{branch cut} starting at the branch~point\begin{equation}\hat s\equiv(m_a+m_b+m_c+m_d)^2\ .\end{equation}\end{Proposition}\noindent The sum of such four-quark branch-cut contributions generates the desired tetraquark~poles.

For a specific (perturbative) QCD contribution to a correlation function (\ref{4}), the question of the presence or absence of a four-quark $s$-channel branch cut can be decided \emph{systematically}: its existence may unambiguously be verified or excluded by consulting, i.e., more frankly,~by solving the relevant set of Landau equations \cite{LDL}. Examples of the application of the Landau equations have been given, also for the flavour-cryptoexotic states (\ref{3F}), by References~\cite{LMS-N3,LMS-N7,LMS-A3}.

When reinstalling, in the four-point correlation function (\ref{4}), the quark-\emph{flavour} indices~of its conventional-meson interpolating currents (\ref{b}), one notices that, with respect to the~\emph{flavour} arrangements in its initial and final states, for the flavour content of any flavour-cryptoexotic tetraquark of the type (\ref{3F}) there exist two different conceivable configurations, highlighted~by\begin{Definition}\label{dPR}An (according to Definition~\ref{dFC}) flavour-cryptoexotic correlation function (\ref{4}) is labelled\begin{itemize}\item\emph{flavour-preserving} \cite{LMS-N3+} if the incoming and outgoing quark-flavour distributions are equal,~but\item\emph{flavour-rearranging} \cite{LMS-N3+} if the incoming and outgoing quark-flavour distributions do not~agree.\end{itemize}\end{Definition}\noindent In principle, in theoretical tetraquark analyses (such as those touched on in Sections~\ref{sN} and \ref{sS}), the necessity of these subdivisions of the correlation functions (\ref{4}) into two disjoint categories has to be taken into account wherever applicable.\footnote{The set of all ``doubly flavoured'' tetraquark mesons, each containing either two quarks or two antiquarks~of~one and the same flavour, i.e., $T=[\overline q_a\,q_b\,\overline q_c\,q_b]$ (with $a\ne b$ and $b\ne c$), or $T=[\overline q_a\,q_b\,\overline q_a\,q_c]$ (with $a\ne b$ and~$a\ne c$),~has been discussed in References~\cite{LMS-N3,LMS-N3.1}: the quark rearrangement of any such state results in the \emph{same} state; hence, the discrimination between flavour-\emph{preserving} and flavour-\emph{rearranging} distribution is neither possible nor~necessary.} Consequently, for the flavour-\emph{cryptoexotic} objects of present desire a split discussion of two variants of correlation functions is in~order:\begin{eqnarray}&\left\langle{\rm T}\!\left(j_{\overline ab}(y)\,j_{\overline bc}(y')\,j^\dag_{\overline ab}(x)\,j^\dag_{\overline bc}(x')\right)\right\rangle\ ,\qquad\left\langle{\rm T}\!\left(j_{\overline ac}(y)\,j_{\overline bb}(y')\,j^\dag_{\overline ac}(x)\,j^\dag_{\overline bb}(x')\right)\right\rangle\ ,&\label{cfp}\\&\left\langle{\rm T}\!\left(j_{\overline ab}(y)\,j_{\overline bc}(y')\,j^\dag_{\overline ac}(x)\,j^\dag_{\overline bb}(x')\right)\right\rangle\ .&\label{cfr}\end{eqnarray}

The decisive move towards a discussion of the flavour-cryptoexotic tetraquarks~(\ref{fce})~that complies with their exotic nature is the identification of those contributions to the four-point correlation functions (\ref{4}) that belong to the tetraquark-phile type demanded by~Proposition~\ref{dpTP}. For the case of the \emph{three-flavour} tetraquarks (\ref{3F}) adopted for illustration, the crucial four-point correlation functions are precisely those of the form (\ref{cfp}) and (\ref{cfr}). The systematic~and~thorough scrutiny for their \mbox{tetraquark-phile} contributions may be accomplished, order by order~in~the (perturbative) expansion of the four-point correlation functions (\ref{4}) with respect to the~strong coupling (\ref{as}), by application of the (analytic) tool provided by the Landau equations \cite{LDL}. For lower-order flavour-preserving cases, this can be achieved by a mere visual inspection: their separability either is obvious or arises from the vanishing trace of all the generators~of~SU(3).

\begin{figure}[htb]\centering
\includegraphics[width=10.566cm]{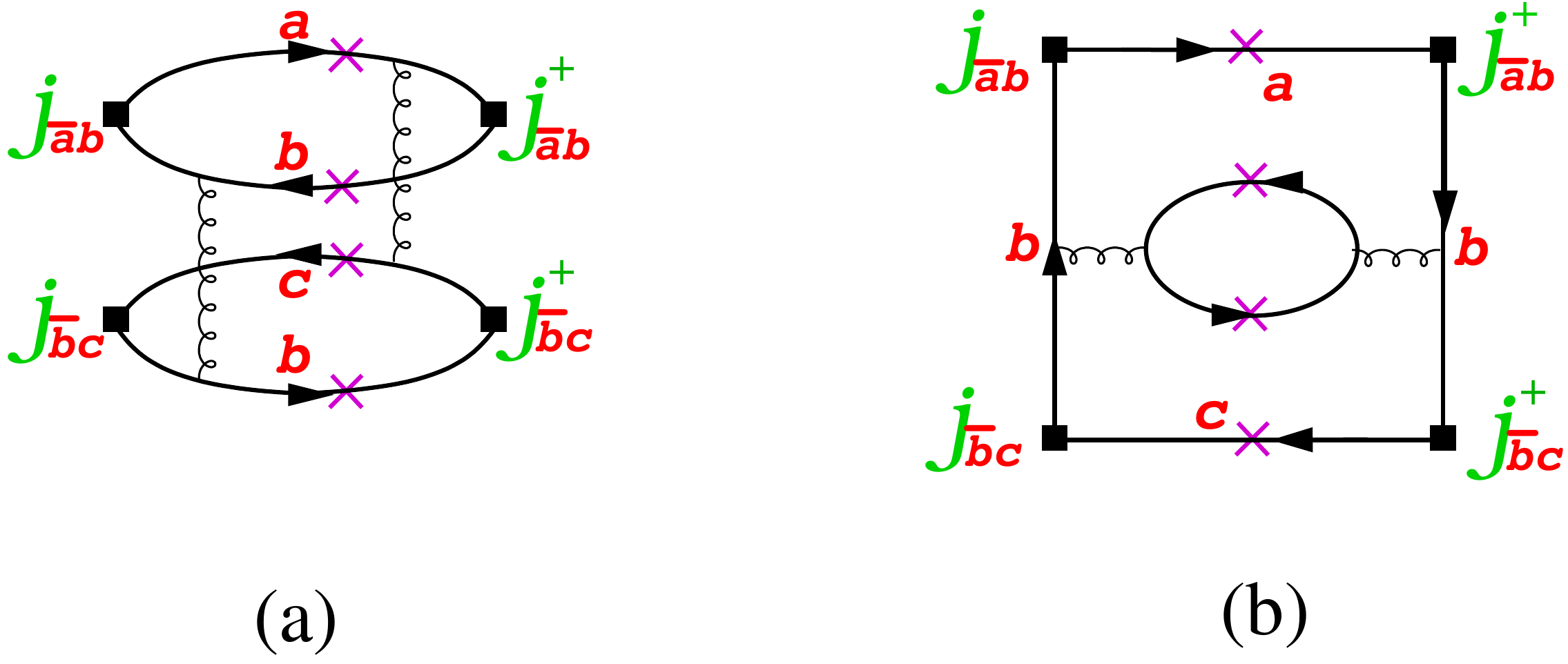}\caption{Flavour-cryptoexotic four-operator correlation function (\ref{4}) of the flavour-preserving~type~(\ref{cfp}), specified in Definition~\ref{dPR}: exemplary contributions \cite{LMS-N1,LMS-N4} established to be tetraquark-phile (according to Proposition~\ref{dpTP}), of the lowest perturbative order capable of satisfying the preconditions of Proposition~\ref{dpTP}, that is, of the order $O(\alpha_{\rm s}^2)$. This order emerges from two internal exchanges of gluons, depicted~in~form of curly lines. The purple crosses highlighting quark propagators identify those four (anti-) quarks~that (by means of branch cuts) \emph{may} contribute to the formation of tetraquark poles. Unlike the contributions of the kind (\textbf{a}), the contributions of the shape (\textbf{b}) do involve one internal quark loop of \emph{arbitrary}~flavour.\label{F4p}}\end{figure}

\begin{figure}[htb]\centering
\includegraphics[width=10.566cm]{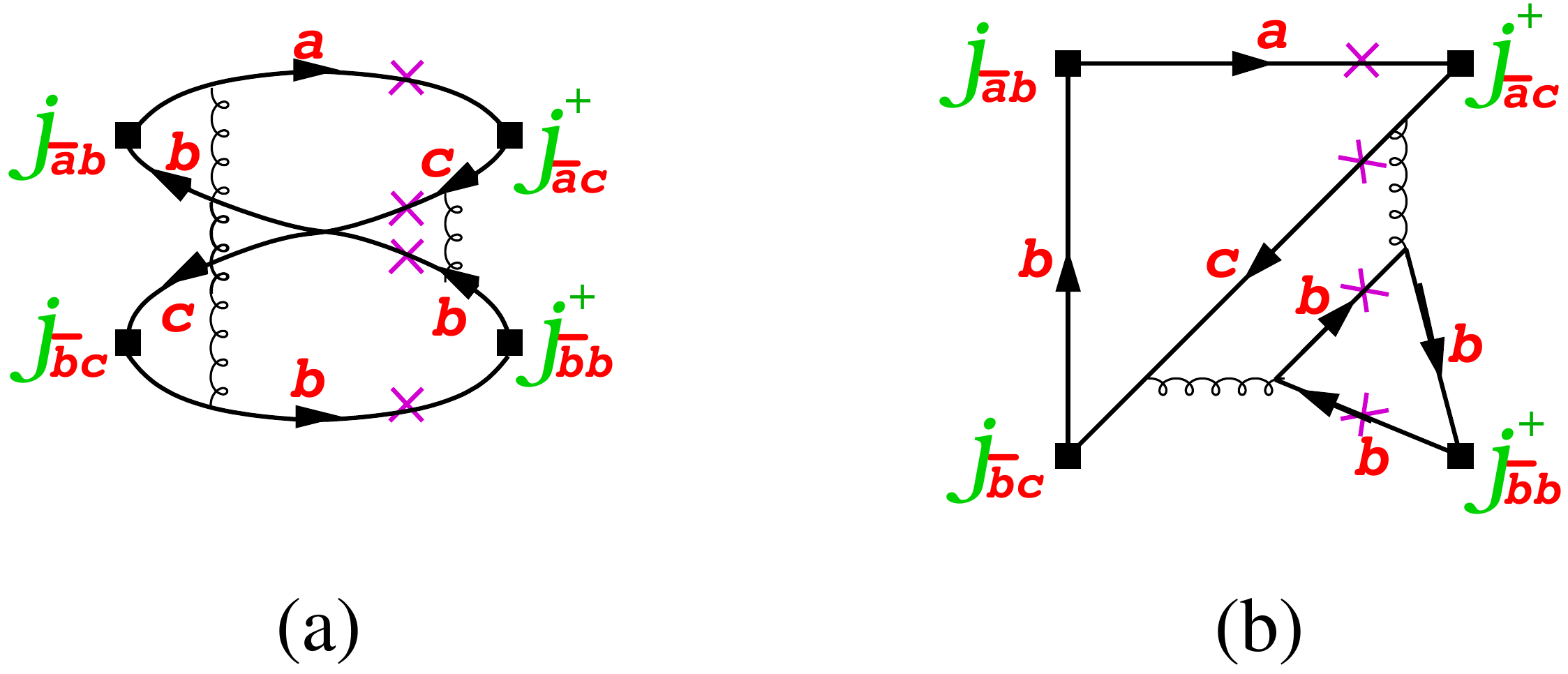}\caption{Flavour-cryptoexotic four-current correlation function (\ref{4}) of the flavour-rearranging type~(\ref{cfr}), specified in Definition~\ref{dPR}: exemplary contributions \cite{LMS-N1,LMS-N4} established to be tetraquark-phile (according to Proposition~\ref{dpTP}), of the lowest perturbative order capable of satisfying the preconditions of Proposition~\ref{dpTP}, that is, of the order $O(\alpha_{\rm s}^2)$. This order emerges from two internal exchanges of gluons, depicted~in~form of curly lines. The purple crosses highlighting quark propagators identify those four (anti-) quarks that (by means of branch cuts) may contribute to the formation of tetraquark poles. The contributions~differ, \emph{decisively} (see Section~\ref{sN}), in the number of closed quark loops: one in the case (\textbf{a}) but two in the~case~(\textbf{b}).\label{F4r}}\end{figure}

For the flavour-cryptoexotic tetraquarks (\ref{3F}), the overall outcome \cite{LMS-N1,LMS-N3} of these studies is that, for \emph{both} the flavour-rearranging quark distribution (\ref{cfr}) and the flavour-preserving~quark distributions (\ref{cfp}), the tetraquark-phile contributions to the four-point correlation function (\ref{4}) cannot show up at a lower than second order of the strong coupling (\ref{as}). In other~words, all of the tetraquark-phile contributions are, necessarily, at least of the order $O(\alpha_{\rm s}^2)$, corresponding to two internal gluon exchanges (of the appropriate topology, of course), or even higher ones. Insights of this kind prove to be of utmost importance for the exploitation of such correlation functions in Sections~\ref{sN} and \ref{sS}. For the two types (\ref{cfp}) and (\ref{cfr}) of flavour-cryptoexotic four-point correlation functions, some tetraquark-phile contributions of lowest order $O(\alpha_{\rm s}^2)$ illustrating their conceivable topology are exemplified, for both varieties of \emph{flavour-preserving} correlation functions (\ref{cfp}), by Figure~\ref{F4p} and, for all \emph{flavour-rearranging} correlation functions (\ref{cfr}), by Figure~\ref{F4r}.

\section{Increase Without Bound of the Number of Colours Entails Useful Qualitative Insights}\label{sN}In order to diminish the obstacles posed by the complexity of QCD to all the attempts~of working out its implications exactly, already almost fifty years ago 't Hooft \cite{GH,GH+} suggested to generalize the gauge group of QCD, SU(3), to SU($N_{\rm c}$), the non-Abelian, special unitary Lie group of degree $N_{\rm c}$. Within the class of gauge theories thus defined, the simplification aimed at may then be achieved by considering the \emph{limit} of the number of colour degrees~of~freedom, $N_{\rm c}$, increasing to infinity, $N_{\rm c}\to\infty$, and the relevant \emph{expansion} thereabout, in powers of $1/N_{\rm c}$. Keeping the large-$N_{\rm c}$ limit under control demands the \emph{product} $N_{\rm c}\,\alpha_{\rm s}$ of the number~of colours and the strong fine-structure coupling (\ref{as}) to remain finite for $N_{\rm c}\to\infty$; this can be assured~by postulating for the strong-interaction strength, that is, for either the strong coupling $g_{\rm s}$ or the strong fine-structure coupling $\alpha_{\rm s}$, its adequate (viz., more frankly, sufficiently rapid)~\emph{decrease}:\begin{align}g_{\rm s}\propto\frac{1}{\sqrt{N_{\rm c}}}&=O(N_{\rm c}^{-1/2})\xrightarrow[N_{\rm c}\to\infty]{}0\label{LNLgv}\\&\,\Updownarrow\nonumber\\\alpha_{\rm s}\propto\frac{1}{N_{\rm c}}&=O(N_{\rm c}^{-1})\xrightarrow[N_{\rm c}\to\infty]{}0\ .\label{LNLav}\end{align}The concomitant generalization of the two types of dynamical degrees of freedom of QCD to their counterparts in any of the quantum field theories that arise for $N_{\rm c}\ne3$ is, unfortunately, just partially unique: Beyond doubt, the gluonic gauge vector bosons \emph{have} to be placed in the $(N_{\rm c}^2-1)$-dimensional adjoint representation of SU($N_{\rm c}$). For any of the quarkonic \emph{fermions} (of given quark flavour $a$), however, the admittedly most common but by no means compulsory \emph{option} is their assignment still to the $N_{\rm c}$-dimensional fundamental representation of SU($N_{\rm c}$).\footnote{The other alternative here would be to attribute the fermionic dynamical degrees of freedom of large-$N_{\rm c}$ QCD to the $\frac{1}{2}\,N_{\rm c}\,(N_{\rm c}-1)$-dimensional antisymmetric representation of ${\rm SU}(N_{\rm c})$: for the standard-QCD case $N_{\rm c}=3$, this antisymmetric representation of ${\rm SU}(N_{\rm c})$ is three-dimensional, just as the fundamental representation of~${\rm SU}(N_{\rm c})$. Unsurprisingly, the predictions of large-$N_{\rm c}$ QCD with its quarks in that antisymmetric representation of~${\rm SU}(N_{\rm c})$ differ significantly from those of large-$N_{\rm c}$ QCD with the quarks in the fundamental representation of ${\rm SU}(N_{\rm c})$ (as recalled in Reference~\cite{CLas}). For uniqueness, it seems advisable to confine the present analysis to a definite~choice.} Hence, the dependence of perturbative large-$N_{\rm c}$ QCD contributions to a correlation function [like the one of the kind (\ref{4})] on the number of colours, $N_{\rm c}$, may be immediately inferred~from the number of involved \emph{closed} colour loops and the number of involved strong couplings~(\ref{as}).

Within large-$N_{\rm c}$ QCD, any colour-singlet conventional mesons $M_{\overline ab}$ (with masses~$M_{M_{\overline ab}}$) continue\footnote{In contrast, for \emph{all} colour-singlet (ordinary) baryons the numbers of quark or antiquark constituents rise~with~$N_{\rm c}$.} to be given by bound states of \emph{one} quark of flavour~$b$ and \emph{one} antiquark of~flavour~$a$:\begin{equation}M_{\overline ab}=[\overline q_a\,q_b]\ ,\qquad a,b\in\{u,d,s,c,b,t(,\dots?)\}\ .\label{cm}\end{equation}

Notationally still skipping all explicit reference to any spin or parity degrees of freedom by dropping all Dirac matrices, polarization vectors, momenta, etc., the nonvanishing~\emph{matrix element} of any conventional-meson interpolating operator (\ref{b}) between the state of the related conventional meson (\ref{cm}) and the vacuum defines the decay constant $f_{M_{\overline ab}}$ of that meson~$M_{\overline ab}$:\begin{equation}f_{M_{\overline ab}}\equiv\langle0|j_{\overline ab}|M_{\overline ab}\rangle\ne0\ .\label{f}\end{equation}Because of the single \emph{quark loop} involved, the large-$N_{\rm c}$ behaviour of $f_{M_{\overline ab}}$ is easy to guess~\cite{EW}:\begin{equation}f_{M_{\overline ab}}\propto\sqrt{N_{\rm c}}=O(N_{\rm c}^{1/2})\ .\label{fN}\end{equation}

Large-$N_{\rm c}$ considerations allow to shed light on crucial aspects of any \emph{flavour-cryptoexotic} tetraquarks (\ref{3F}): their total decay width and their potential mixing with conventional~mesons. In these analyses, those contributions to any flavour-cryptoexotic correlation functions~(\ref{cfp})~or (\ref{cfr}) that (by Proposition~\ref{dpTP}) may be expected to be \textbf{t}etraquark-\textbf{p}hile and thus might take~part~in an eventual formation of some \textbf{t}etraquark \textbf{p}oles will be identified by the subscript label~``tp''.

\subsection{Total Decay Width of Flavour-Cryptoexotic Tetraquarks: Upper Bounds on Large-$N_{\rm c}$ Behaviour}\label{w}The experimental observability in nature of some tetraquark meson requires a sufficient narrowness of that hadron: Compared to its mass, its decay width $\Gamma$ should not be too~broad. For the large-$N_{\rm c}$ generalization of QCD, supposing the tetraquark mass to be independent of the number $N_{\rm c}$ of colours, the decay width of such a tetraquark should not \emph{grow} with~$N_{\rm c}$~\cite{SW}. Therefore, the exploration of the large-$N_{\rm c}$ behaviour of the tetraquark decay widths seems to be in order. Tetraquark states \emph{need not} show up already at largest \emph{tetraquark-phile} orders~of~$N_{\rm c}$: generally, merely upper bounds on any extracted large-$N_{\rm c}$ dependences should be expected.

Now, the decay of some tetraquark meson state, $T$, may be assumed to be dominated by the latter's decay into a pair of conventional mesons of the generic kind (\ref{cm}). Hence, the total decay \emph{width} of $T$ will be governed by the corresponding amplitudes $A(T\longleftrightarrow M_{\overline ab}\,M_{\overline cd})$ that encode all transitions between the tetraquark $T$ and a conventional-meson pair $M_{\overline ab}$~and~$M_{\overline cd}$. These tetraquark--two-conventional-meson \emph{transition} amplitudes $A$, in turn, can be extracted from the intermediate-state tetraquark-pole contributions to the \emph{scattering} amplitudes for the appropriate scattering processes of two conventional mesons into two conventional~mesons.

Specifically, zooming in to the \emph{flavour-cryptoexotic} states (\ref{3F}), the leading-$N_{\rm c}$ behaviour of \emph{all tetraquark-phile} contributions to the correlation functions (\ref{cfp}) and (\ref{cfr}) is easily pinned~down.\begin{itemize}\item In the flavour-preserving case (\ref{cfp}), both sorts of \emph{tetraquark-phile} QCD-level contributions of lowest perturbative order illustrated in Figure~\ref{F4p}a and Figure~\ref{F4p}b exhibit rather~similar characteristics: both types are built from two closed quark loops and two internal gluon exchanges; this then translates into \emph{two} closed \emph{colour} loops and \emph{two} powers of the strong coupling (\ref{LNLav}). The order of all $N_{\rm c}$-\emph{leading} contributions thus is $O(N_{\rm c}^2\,\alpha_{\rm s}^2)=O(N_{\rm c}^0)$~\cite{LMS-N1,LMS-N3}.\item In the flavour-rearranging case (\ref{cfr}), the two examples of \emph{tetraquark-phile} contributions of lowest perturbative-QCD order depicted in Figure~\ref{F4r}a and Figure~\ref{F4r}b are of undoubtedly unlike structures: The contributions exemplified in Figure~\ref{F4r}a involve merely one~closed quark loop and two internal gluon exchanges. This corresponds to a \emph{single} closed~\emph{colour} loop and \emph{two} powers of the strong coupling (\ref{LNLav}). On the other hand, any contribution of the sort shown in Figure~\ref{F4r}b is formed by two closed quark loops and two internal gluon exchanges, which is tantamount to \emph{two} closed \emph{colour} loops and \emph{two} powers of the strong coupling (\ref{LNLav}). This entails a large-$N_{\rm c}$ dependence of the order $O(N_{\rm c}^2\,\alpha_{\rm s}^2)=O(N_{\rm c}^0)$~\cite{LMS-N1,LMS-N3}.\end{itemize}In total, the upper bounds on the large-$N_{\rm c}$ behaviour of the tetraquark-phile contributions to both types of the flavour-cryptoexotic correlation functions (\ref{cfp}) and (\ref{cfr}) prove to be~identical:\footnote{In contrast to this, in the class of tetraquark mesons that exhibit the maximum number of four mutually~different open quark flavours (reviewed in Ref.~\cite{L:MT1}), for flavour-preserving correlation functions (\ref{4}), on the one hand,~and related flavour-rearranging correlation functions (\ref{4}), on the other hand, the large-$N_{\rm c}$ behaviour of the respective tetraquark-phile contributions \emph{differs} by one order of $N_{\rm c}$. This discrepancy may be accommodated, or dealt with, by postulating, or enabling, the pairwise occurrence of the particular tetraquark species under consideration~\cite{LMS-N1}.}\begin{align}&
\left\langle{\rm T}\!\left(j_{\overline ab}(y)\,j_{\overline bc}(y')\,j^\dag_{\overline ab}(x)\,j^\dag_{\overline bc}(x')\right)\right\rangle_{\rm tp}=O(N_{\rm c}^2\,\alpha_{\rm s}^2)=O(N_{\rm c}^0)\ ,\label{fpt1}\\[.23ex]&
\left\langle{\rm T}\!\left(j_{\overline ac}(y)\,j_{\overline bb}(y')\,j^\dag_{\overline ac}(x)\,j^\dag_{\overline bb}(x')\right)\right\rangle_{\rm tp}=O(N_{\rm c}^2\,\alpha_{\rm s}^2)=O(N_{\rm c}^0)\ ,\label{fpt2}\\[.23ex]&
\left\langle{\rm T}\!\left(j_{\overline ab}(y)\,j_{\overline bc}(y')\,j^\dag_{\overline ac}(x)\,j^\dag_{\overline bb}(x')\right)\right\rangle_{\rm tp}=O(N_{\rm c}^2\,\alpha_{\rm s}^2)=O(N_{\rm c}^0)\ .\label{frt}\end{align}

Recalling that the finding (\ref{fN}) for the conventional-meson decay constants (\ref{f}) enters in any four-point correlation functions (\ref{4}) with fourth power, the large-$N_{\rm c}$ dependences of both encountered tetraquark--two-conventional-meson \emph{transition} amplitudes $A(T\longleftrightarrow M_{\overline ab}\,M_{\overline bc})$ and $A(T\longleftrightarrow M_{\overline ac}\,M_{\overline bb})$, as deduced from the findings (\ref{fpt1}), (\ref{fpt2}), and (\ref{frt}), are parametrically identical: \emph{both} are of the order $O(N_{\rm c}^{-1})$. More precisely, these transition amplitudes exhibit a \emph{parametrically} identical decrease proportional to $1/N_{\rm c}$ with increasing number $N_{\rm c}$ of colours. The associated tetraquark decay width $\Gamma$ involves the {squares} of these transition~amplitudes. Hence, the decay width $\Gamma(T)$ of all \emph{flavour-cryptoexotic} tetraquarks (\ref{3F}) is of the order $O(N_{\rm c}^{-2})$. Schematically, the relations among these quantities are perhaps best summarized in the~form\begin{equation}\underbrace{A(T\longleftrightarrow M_{\overline ab}\,M_{\overline bc})=O(N_{\rm c}^{-1})\quad\stackrel{\fbox{$N_{\rm c}$ order}}{\mbox{\boldmath$\equiv$}}\quad A(T\longleftrightarrow M_{\overline ac}\,M_{\overline bb})=O(N_{\rm c}^{-1})}_{\mbox{$\Longrightarrow\qquad\Gamma(T)=O(N_{\rm c}^{-2})$}}\ .\label{cpo}\end{equation}

In Table~\ref{W}, the large-$N_{\rm c}$ \emph{upper bound} (\ref{cpo}) found for the decay widths $\Gamma$ is confronted with the outcomes of comparable considerations from (marginally) different perspectives~\cite{KP,MPR}.\begin{table}[H]\caption{Theoretically predicted upper bounds on the large-$N_{\rm c}$ behaviour of the total decay widths~$\Gamma$~of the flavour-cryptoexotic tetraquarks (\ref{fce}). Distilled by condensing Table~2 of Ref.~\cite{LMS-N4} or Table~3 of~Ref.~\cite{L:MT1}.\label{W}}\begin{center}\begin{tabular}{lcr}\toprule\textbf{Author Collective}&\textbf{Large-$N_{\rm c}$ Total Decay Width}&\textbf{References}\\\midrule Knecht and Peris&$O(1/N_{\rm c})$&\cite{KP}\\Maiani, Polosa, and Riquer&$O(1/N_{\rm c}^3)$&\cite{MPR}\\Lucha, Melikhov, and Sazdjian&$O(1/N_{\rm c}^2)$&\cite{LMS-N1,LMS-N3}\\\bottomrule\end{tabular}\end{center}\end{table}\noindent The undoubtedly conspicuous \emph{incongruences} of the large-$N_{\rm c}$ expectations compiled in~Table~\ref{W} originate very likely either from the --- from the point of view of Proposition~\ref{dpTP} \emph{unjustified} and thus perhaps misleading --- trust \cite{KP} in contributions that do not qualify as tetraquark-phile or from the neglect \cite{MPR} of the $N_{\rm c}$-leading contributions of the kind put forward~in~Figure~\ref{F4r}b.

The lessons to be learned from the bulk of all large-$N_{\rm c}$ considerations of this kind~can be subsumed as follows: The total \emph{decay widths}, $\Gamma(T)$, of the flavour-cryptoexotic tetraquarks~(\ref{3F}) neither grow with $N_{\rm c}$ nor are independent of $N_{\rm c}$. Rather, they decrease with $N_{\rm c}$. This implies that, from the large-$N_{\rm c}$ aspect, the flavour-cryptoexotic tetraquarks are narrow~\mbox{hadrons. The} comparison with the large-$N_{\rm c}$ behaviour of generic conventional mesons (\ref{cm}) reveals,~maybe surprisingly, differences. In the large-$N_{\rm c}$ limit, the total decay widths of flavour-cryptoexotic tetraquarks (\ref{3F}), in Equation~(\ref{cpo}) found to be (at most) of the order $O(N_{\rm c}^{-2})$, decrease~(at~least) proportional to $1/N_{\rm c}^2$ and, hence, \emph{definitely} faster than the conventional-meson decay~widths, which --- having been found \cite{EW} to be of the order $O(N_{\rm c}^{-1})$ --- decrease proportional~to~$1/N_{\rm c}$. \emph{Flavour-cryptoexotic} tetraquarks~(\ref{3F}) are \emph{parametrically} narrower than the conventional mesons.

\subsection{Mixing of Flavour-Cryptoexotic Tetraquark Mesons and Conventional Mesons: Large-$N_{\rm c}$ Limit}As already pointed out in Section~\ref{sE}, any flavour-\emph{cryptoexotic} multiquark hadron state,~in Definition~\ref{dFC} characterized by presenting a set of \emph{discernible} quark-flavour degrees of freedom which is identical to the flavour content of a conventional hadron state, may undergo mixing with such a conventional hadron state if not forbidden by any conserved degrees of freedom. In particular, \emph{any} flavour-cryptoexotic tetraquark (\ref{fce}), evidently involving the two \emph{open} quark flavours $\overline q_a$ and $q_c$, has to be supposed to mix with any conventional meson (\ref{cm}) composed of the matching quark flavours, $M_{\overline ac}$. In view of this, within large-$N_{\rm c}$ considerations one urgent question immediately emerges: What is the impact of these tetraquark--conventional-meson mixings on the large-$N_{\rm c}$ behaviour of the flavour-cryptoexotic tetraquarks (\ref{fce})? Does~this sort of mixing \emph{qualitatively} modify any large-$N_{\rm c}$ predictions for flavour-cryptoexotic tetraquarks?

The strength of any coupling of a flavour-cryptoexotic tetraquark $T$ of quark content (\ref{fce}) and an adequate conventional meson $M_{\overline ac}$ may be encoded by their mixing parameter $g_{TM_{\overline ac}}$. The behaviour of the mixing parameters $g_{TM_{\overline ac}}$ in the large-$N_{\rm c}$ limit can, without problem, be inferred from the knowledge of the \emph{generic} large-$N_{\rm c}$ dependences of the conventional-meson decay constants $f_{M_{\overline ab}}$ recalled by Eq.~(\ref{fN}), of the two available types of amplitudes~describing transitions between a tetraquark and a pair of conventional mesons, found in Eq.~(\ref{cpo}),~and of the amplitudes $A(M_{\overline ac}\longleftrightarrow M_{\overline ab}\,M_{\overline bc})$ of all couplings among three \emph{conventional} mesons~\cite{EW},\begin{equation}A(M_{\overline ac}\longleftrightarrow M_{\overline ab}\,M_{\overline bc})\propto\frac{1}{\sqrt{N_{\rm c}}}=O(N_{\rm c}^{-1/2})\xrightarrow[N_{\rm c}\to\infty]{}0\ .\label{3m}\end{equation}

For instance, upon truncation or neglect of all quantities regarded as independent of $N_{\rm c}$ (such as any propagators of internal and external mesons) the portion of the tetraquark-phile contribution (\ref{fpt1}) to one of the flavour-cryptoexotic correlation functions (\ref{cfp}) that~receives just a single tetraquark--conventional-meson mixing is boiled down to the large-$N_{\rm c}$ constraint~\cite{LMS-N1}\begin{equation}f_M^4\,A(T\longleftrightarrow M_{\overline ab}\,M_{\overline bc})\,g_{TM_{\overline ac}}\,A(M_{\overline ac}\longleftrightarrow M_{\overline ab}\,M_{\overline bc})=O(N_{\rm c}^0)\ .\end{equation}Similar studies can be performed for tetraquark-phile contributions exhibiting \emph{more} than~one such mixing \cite{LMS-N3}. The large-$N_{\rm c}$ \emph{upper bounds} resulting thereof ``fortunately'' agree. The~mixing parameters $g_{TM_{\overline ac}}$ are of the order $O(N_{\rm c}^{-1/2})$ and, thus, decrease not \emph{slower} than $1/\sqrt{N_{\rm c}}$~\cite{LMS-N1,LMS-N3}:\begin{equation}g_{TM_{\overline ac}}\propto\frac{1}{\sqrt{N_{\rm c}}}=O(N_{\rm c}^{-1/2})\xrightarrow[N_{\rm c}\to\infty]{}0\ .\label{mp}\end{equation}All \emph{products} of one three-meson amplitude (\ref{3m}) and one mixing parameter (\ref{mp}) reproduce the large-$N_{\rm c}$ behaviour of the transition amplitudes (\ref{cpo}); thus, the mixing of a tetraquark~with its conventional-meson quark-flavour counterparts will not alter the findings of Subsection~\ref{w}:\begin{equation}A(T\longleftrightarrow M_{\overline ab}\,M_{\overline bc})=O(N_{\rm c}^{-1})\hspace{1.8777ex}{\widehat=}\hspace{1.8777ex}g_{TM_{\overline ac}}\,A(M_{\overline ac}\longleftrightarrow M_{\overline ab}\,M_{\overline bc})=O(N_{\rm c}^{-1})\ .\label{=}\end{equation}

\section{Application of the QCD Sum-Rule Formalism to Multiquarks: Immediate Implication}\label{sS}Ideally (or, perhaps, from a somewhat \emph{puristic} point of view), the theoretical description of bound states in terms of some underlying quantum field theory should be carried out~in~a manner that is both analytical and nonperturbative. For the strong interactions, an approach that comes pretty close to these two requirements is the framework of \emph{QCD~sum rules} \cite{SVZ,RRY,CK}. The basic idea behind the latter tool is the construction of relations between the fundamental parameters of QCD, on the one hand, and (experimentally observable)~properties of hadrons emerging as bound states of the set of \emph{dynamical} QCD degrees of freedom, on the other~hand.

This derivation can be enacted or achieved by, first, the definition of suitable correlation functions that involve hadron interpolating operators, related (among others) to the hadrons of interest, followed by the simultaneous \emph{evaluation} of one and the same correlation function, at both the phenomenological level of the hadronic states and the fundamental level of~QCD:\begin{itemize}\item At the hadron level, the insertion of a complete set of hadron states brings into the game all hadrons potentially contributing in form of intermediate states (more precisely, their observable characteristics, such as masses, decay constants, transition~amplitudes,~etc.); among the latter hadrons, there should show up the particular multiquark under study.\item At the QCD level, the conversion \cite{KGW} of the nonlocal product of interpolating operators in any such correlation function into a series of local operators enables the separation of the perturbative from the nonperturbative contributions: The \emph{perturbative} contributions might be obtained, for lower orders of the strong coupling, order by order (as discussed in Section~\ref{sC}). The \emph{nonperturbative} contributions, however, cannot be derived (at present) from the underlying quantum field theory. They can be parametrized by quantities that may be inferred from experiment and can be interpreted as \emph{effective} parameters~of~QCD.\end{itemize}Equating the eventual outcomes of the two procedures generates the desired QCD sum~rules.

When applying the generally valid QCD sum-rule technique, specifically, to any type of \emph{multiquark state}, particular attention must be paid to two ``aggravating inconveniences'': the construction of (suitable) \emph{multiquark} interpolating operators and, in view of the discussion in Section~\ref{sC}, the multiquark adequacy \cite{LMS-A1,LMS-A3} of the predictions emerging from this technique.\footnote{In the case of \emph{pentaquark} states, that is, \emph{baryonic} multiquark hadrons built up from four quarks and one~antiquark, problems of similar nature have been noted (and tackled along a somewhat different path) in References~\cite{PQ1,PQ2}.}\begin{enumerate}\item If narrowing down the envisaged quest for multiquark-adequate QCD sum rules to the subcategory of multiquark exotics that is formed by all tetraquark mesons, the problem of identifying, for particular states, the most appropriate set of tetraquark interpolating operators is considerably mitigated by the observation that, upon application of proper Fierz transformations \cite{MF}, every \emph{colour-singlet} operator constructed of two quark fields and two antiquark fields can easily be \emph{rearranged} to a linear combination of products~of two \emph{conventional-meson} interpolating operators (\ref{b}). As far as the quark~\emph{flavour} quantum numbers $a,b,c,d$ are concerned, not more than two products of such kind are available:\begin{equation}j_{\overline ab}(x)\,j_{\overline cd}(x)\ ,\qquad j_{\overline ad}(x)\,j_{\overline cb}(x)\ .\label{t}\end{equation}The set (\ref{t}) of products of colour-singlet quark--antiquark bilinear operators (\ref{b}) may~be regarded to provide a sort of basis of the space of all tetraquark interpolating operators.\item The \emph{product} nature of an element of the tetraquark interpolating operator basis (\ref{t}) may be imagined to arise from the identification or ``contraction'' of the configuration-space coordinates of appropriate pairs of quark-bilinear currents (\ref{b}) like those showing up in each of the four-point correlation functions (\ref{4}). This fact, in turn, offers the opportunity to construct correlation functions that involve either a single or even a pair of \emph{tetraquark} interpolating operators by subjecting appropriately selected correlation functions (\ref{4}) of four quark-bilinear operators (\ref{b}) to one or two of these spatial-coordinate contractions.\item In the course of invoking the standard QCD sum-rule technique for the investigation of multiquarks, this tool's intended improvement, dubbed its \emph{multiquark adequacy}, may be accomplished by diminishing, to the utmost reasonable extent, all its ``contaminations'' by contributions evidentially irrelevant to any exotic state momentarily under scrutiny. For the tetraquark mesons, this demands to retain exclusively QCD-level contributions to correlation functions that are tetraquark-phile, in full compliance with Proposition~\ref{dpTP}, and to carefully \emph{match} any of these contributions with the corresponding mirror~images in the set of hadron-level contributions to the very same kind of correlation functions.\footnote{For tetraquarks displaying four different quark flavours (chosen just for simplicity), this \emph{modus operandi}~has~been proposed and analyzed, in \emph{great} detail, in References \cite{LMS-A1,LMS-A3} and recently revisited, in \emph{due} detail, in Reference~\cite{L:MT1}.}\item Focusing one very last time to the subset of all flavour-cryptoexotic tetraquarks~(\ref{3F})~that involve three disparate quark flavours, the analysis carried out in Section~\ref{sC} implies that tetraquark-phile contributions to any correlation function that is on the verge of getting (calculationally) converted to a QCD sum rule cannot arise before the \emph{second order} of the perturbative expansion in powers of the strong coupling strength $\alpha_{\rm s}$. All (concomitant) contributions at the hadronic level, however, ought to be thoroughly disentangled with respect to their actual \emph{relevance} for each tetraquark state considered. This task~proves to be (comparatively) straightforward for all flavour-preserving~correlation functions~(\ref{cfp}). For any flavour-rearranging correlation function (\ref{cfr}), the case-by-case judgement~might turn out to be in order. The actual \emph{feasibility} of any such analysis has been claimed and a \emph{conceivable} route briefly indicated, for the flavour-preserving quark distributions (\ref{cfp}), in Reference~\cite{LMS-A1} but, for the flavour-rearranging quark distribution (\ref{cfr}), in Reference~\cite{LMS-A3}.\end{enumerate}The \emph{tetraquark-adequate} QCD sum rules gained from this optimization effort may be expected to provide, for various basic properties of \emph{any} flavour-cryptoexotic tetraquark (\ref{3F}) considered, predictions of inevitably higher credibility. Among these characteristics of any such state~are its mass, its decay constants, given by the vacuum--tetraquark matrix elements of the~hadron interpolating operators (\ref{t}), and all strengths of its couplings to two quark-bilinear currents, given by the vacuum--tetraquark matrix elements of not contracted pairs of the operators~(\ref{b}).

\section{Conclusions: Promising Prospects of Approaches to Flavour-Cryptoexotic Tetraquarks}The possible existence of multiquark exotic hadrons in form of tightly bound states, as a hardly evitable implication of the strong interactions represented by QCD, has been~\emph{predicted} already long ago. Substantial \emph{experimental evidence} for the actual existence, and observability, of multiquark hadrons has been accumulated only comparatively recently. The by far largest number of experimental candidates for multiquarks belongs to the class of tetraquarks~of~the \emph{flavour-cryptoexotic} kind: tetraquarks that, by including a quark and an antiquark of the same quark flavour, exhibit, at most, only two \emph{openly} discernible quark flavour quantum~numbers.

For their theoretical interpretations, these experimental findings still form considerable challenges: they constitute, at present, one of the greatest riddles of hadron phenomenology. In order to promote, to some extent, the progress in the \emph{systematic} theoretical investigation of the tetraquark mesons, a rigorous constraint on any meaningful (thus acceptable) theoretical input to analyses of this kind has been proposed \cite{LMS-N1,LMS-N3}. (It goes without saying that each~such type of constraint may merely serve as a necessary but not sufficient condition for the~factual participation of a given tetraquark state in any physical process under study.) Among~others, this tool can then be used to confirm that the decay widths of any tetraquark states should be expected to be parametrically narrower than those of ordinary (quark--antiquark) mesons, or to point out directions towards a more adequate description of tetraquark mesons, by means of suitably adapted QCD sum rules, and has yet to be utilized in multiquark~studies~\cite{TAQSR1+,TAQSR2+,TAQSR3+,TAQSR5+,PCB,C+,TAQSR7+,TAQSR8}.

\vspace{6pt}

\funding{This research received no external funding.}
\dataavailability{Data sharing not applicable.}
\acknowledgments{The author would like to thank both Dmitri I.\ Melikhov and Hagop Sazdjian, for a particularly pleasurable, enjoyable, and inspiring collaboration on various of the topics covered~above.}
\conflictsofinterest{The author declares no conflict of interest.}
\abbreviations{Abbreviations}{The following abbreviation is used in this manuscript:\\

\noindent\begin{tabular}{@{}ll}QCD&quantum chromodynamics\end{tabular}}

\begin{adjustwidth}{-\extralength}{0cm}

\reftitle{References}

\PublishersNote{}
\end{adjustwidth}

\begin{thebibliography}{99}
\bibitem[PDG(22)]{PDG}
Workman, R.L.; et al.
Review of particle physics (2022).
{\em Prog.~Theor.~Exp.~Phys.} {\bf 2022}, {\em 2022}, 083C01.
\bibitem[Lucha(17a)]{LMS-N1}
Lucha, W.; Melikhov, D.; Sazdjian, H.
Narrow exotic tetraquark mesons in large-$N_{\rm c}$ QCD.
{\em Phys.~Rev.~D} {\bf 2017}, {\em 96}, 014022.
\bibitem[Lucha(18b)]{LMS-N2}
Lucha, W.; Melikhov, D.; Sazdjian, H.
Exotic states and their properties from large-$N_{\rm c}$ QCD.
{\em PoS} {\bf 2018}, {\em EPS-HEP 2017}, 390.
\bibitem[Lucha(17b)]{LMS-N3}
Lucha, W.; Melikhov, D.; Sazdjian, H.
Tetraquark and two-meson states at large $N_{\rm c}$.
{\em Eur.~Phys.~J.~C} {\bf 2017}, {\em 77}, 866.
\bibitem[Lucha(18d)]{LMS-N3+}
Lucha, W.; Melikhov, D.; Sazdjian, H.
Constraints from the $1/N_{\rm c}$ expansion on properties of exotic tetraquark mesons.
{\em PoS} {\bf 2018}, {\em Hadron2017}, 233.
\bibitem[Lucha(18x)]{LMS-N3.1}
Lucha, W.; Melikhov, D.; Sazdjian, H.
Narrow-width tetraquarks in large-$N_{\rm c}$ QCD.
{\em EPJ Web Conf.} {\bf 2018}, {\em 182}, 02111.
\bibitem[Lucha(18a)]{LMS-N4}
Lucha, W.; Melikhov, D.; Sazdjian, H.
Exotic tetraquark mesons in large-$N_{\rm c}$ limit: an unexpected great surprise.
{\em EPJ Web Conf.} {\bf 2018}, {\em 192}, 00044.
\bibitem[Lucha(21b)]{LMS-N7}
Lucha, W.; Melikhov, D.; Sazdjian, H.
Tetraquarks in large-$N_{\rm c}$ QCD.
{\em Prog.~Part.~Nucl.~Phys.} {\bf 2021}, {\em 120}, 103867.
\bibitem[Lucha(23)]{L:MT1}
Lucha, W.
Mission target: exotic multiquark hadrons --- sharpened blades.
{\em Universe} {\bf 2023}, {\em 9}, 171.
\bibitem[Lucha(18c)]{LMS-N5}
Lucha, W.; Melikhov, D.; Sazdjian, H.
Are there narrow flavour-exotic tetraquarks in large-$N_{\rm c}$ QCD?.
{\em Phys.~Rev.~D} {\bf 2018}, {\em 98}, 094011.
\bibitem[Landau(59)]{LDL}
Landau, L.D.
On analytic properties of vertex parts in quantum field theory.
{\em Nucl.~Phys.} {\bf 1959}, {\em 13}, 181.
\bibitem[Lucha(19b)]{LMS-A3}
Lucha, W.; Melikhov, D.; Sazdjian, H.
Tetraquark-adequate QCD sum rules for quark-exchange processes.
{\em Phys.~Rev.~D} {\bf 2019}, {\em 100}, 074029.
\bibitem['t Hooft(74)]{GH}
't Hooft, G.
A planar diagram theory for strong interactions.
{\em Nucl.~Phys.~B} {\bf 1974}, {\em 72}, 461.
\bibitem['t Hooft(74a)]{GH+}
't Hooft, G.
A two-dimensional model for mesons.
{\em Nucl.~Phys.~B} {\bf 1974}, {\em 75}, 461.
\bibitem[Cohen(14as)]{CLas}
Cohen, T.D.; Lebed, R.F.
Tetraquarks with exotic flavor quantum numbers at large $N_{\rm c}$ in QCD(AS).
{\em Phys.~Rev.~D} {\bf 2014}, {\em 89}, 054018.
\bibitem[Witten(79)]{EW}
Witten, E.
Baryons in the $1/N$ expansion.
{\em Nucl.~Phys.~B} {\bf 1979}, {\em 160}, 57.
\bibitem[Weinberg(1)]{SW}
Weinberg, S.
Tetraquark mesons in large-$N$ quantum chromodynamics.
{\em Phys.~Rev.~Lett.} {\bf 2013}, {\em 110}, 261601.
\bibitem[Knecht(13)]{KP}
Knecht, M.; Peris, S.
Narrow tetraquarks at large $N$.
{\em Phys.~Rev.~D} {\bf 2013}, {\em 88}, 036016.
\bibitem[Maiani(16)]{MPR}
Maiani, L.; Polosa, A.D.; Riquer, V.
Tetraquarks in the $1/N$ expansion and meson--meson resonances.
{\em J.~High Energy Phys.} {\bf 2016}, {\em 06}, 160.
\bibitem[Shifman(79)]{SVZ}
Shifman, M.A.; Vainshtein, A.I.; Zakharov, V.I.
QCD and resonance physics. Theoretical foundations.
{\em Nucl.~Phys.~B} {\bf 1979}, {\em 147}, 385.
\bibitem[Reinders(85)]{RRY}
Reinders, L.J.; Rubinstein, H.; Yazaki, S.
Hadron properties from QCD sum rules.
{\em Phys.~Reports} {\bf 1985}, {\em 127}, 1.
\bibitem[Colangelo(00)]{CK}
Colangelo, P.; Khodjamirian, A.
QCD sum rules, a modern perspective.
In {\em At the Frontier of Particle Physics --- Handbook of QCD. Boris Ioffe Festschrift}; Shifman, M., Ed.; World Scientific: Singapore, 2001; Vol.~3, p.~1495.
\bibitem[Wilson(69)]{KGW}
Wilson, K.G.
Non-Lagrangian models of current algebra.
{\em Phys.~Rev.} {\bf 1969}, {\em 179}, 1499.
\bibitem[Lucha(19a)]{LMS-A1}
Lucha, W.; Melikhov, D.; Sazdjian, H.
Tetraquark-adequate formulation of QCD sum rules.
{\em Phys.~Rev.~D} {\bf 2019}, {\em 100}, 014010.
\bibitem[Kondo(05)]{PQ1}
Kondo, Y.; Morimatsu, O.; Nishikawa, T.
Two-hadron-irreducible QCD sum rule for pentaquark baryon.
{\em Phys.~Lett.~B} {\bf 2005}, {\em 611}, 93.
\bibitem[Nishikawa(07)]{PQ2}
Nishikawa, T.; Kondo, Y.; Morimatsu, O.; Kanada-En'yo, Y.
Pentaquarks in QCD sum rules.
{\em Prog.~Theor.~Phys.~Suppl.} {\bf 2007}, {\em 168}, 54.
\bibitem[Fierz(37)]{MF}
Fierz, M.
Zur Fermischen Theorie des $\beta$-Zerfalls.
{\em Z.~Phys.} {\bf 1937}, {\em 104}, 553.
\bibitem[LCJC(22)]{TAQSR1+}
Chen, H.-X.; Chen, W.; Zhu, S.-L.
Possible interpretations of the $P_c(4312)$, $P_c(4440)$, and $P_c(4457)$.
{\em Phys.~Rev.~D} {\bf 2019}, {\em 100}, 051501(R).
\bibitem[Pimikov(20)]{TAQSR2+}
Pimikov, A.; Lee, H.-J.; Zhang, P.
Hidden-charm pentaquarks with color-octet substructure in QCD sum rules.
{\em Phys.~Rev.~D} {\bf 2020}, {\em 101}, 014002.
\bibitem[Nora(20)]{TAQSR3+}
Brambilla, N.; Eidelman, S.; Hanhart, C.; Nefediev, A.; Shen, C.-P.; Thomas, C.E.; Vairo, A.; Yuan, C.-Z.
The \emph{XYZ} states: experimental and theoretical status and perspectives.
{\em Phys.~Reports} {\bf 2020}, {\em 873}, 1.
\bibitem[LCJC(22)]{TAQSR5+}
Li, S.-H.; Chen, Z.-S.; Jin, H.-Y.; Chen, W.
Mass of $1^{-+}$ four-quark--hybrid mixed states.
{\em Phys.~Rev.~D} {\bf 2022}, {\em 105}, 054030.
\bibitem[Pal(23)]{PCB}
Pal, S.; Chakrabarti, B.; Bhattacharya, A.
A theoretical investigation on the spectroscopy and structure of the exotic tetraquark~states.
{\em Nucl.~Phys.~A} {\bf 2023}, {\em 1029}, 122559.
\bibitem[Hanhart(22)]{C+}
Hanhart, C.; Nefediev, A.
Do near-threshold molecular states mix with neighboring $\overline QQ$ states?.
{\em Phys.~Rev.~D} {\bf 2022}, {\em 106}, 114003.
\bibitem[Sundu(23)]{TAQSR7+}
Sundu, H.; Agaev, S.S.; Azizi, K.
Axial-vector and pseudoscalar tetraquarks $[ud][\overline c\overline s]$.
{\em Eur.~Phys.~J.~C} {\bf 2023}, {\em 83}, 198.
\bibitem[Dong(23)]{TAQSR8}
Dong, R.-R.; Su, N.; Chen, H.-X.; Cui, E.-L.
QCD sum rule study on the fully strange tetraquark states of $J^{PC}=2^{++}$.
{\em Front.~Phys.} {\bf 2023}, {\em 11}, 1184103.
\end{thebibliography}
\end{document}